\documentclass[%
 aip,
 reprint,
 superscriptaddress,
 amsmath,amssymb,
 floatfix,
]{revtex4-1}

\usepackage{xcolor} 
\usepackage{graphicx}
\graphicspath{{Figures/}}
\usepackage{dcolumn}
\usepackage{bm}
\usepackage{cmap}
\usepackage[utf8]{inputenc}
\usepackage[T1]{fontenc}
\usepackage{textcomp}
\usepackage{mathptmx}
\usepackage{etoolbox}
\usepackage{siunitx}
\usepackage[hidelinks]{hyperref}

\begin{document}

\title[Heteroepitaxial Growth of PbSe on InP via III--V Buffer Layers]{Heteroepitaxial Growth of PbSe on InP via III--V Buffer Layers}

\author{Biridiana Rodriguez}
\affiliation{CREOL, The College of Optics and Photonics, University of
Central Florida, Orlando, FL 32816, USA}

\author{Mark Martino}
\affiliation{CREOL, The College of Optics and Photonics, University of
Central Florida, Orlando, FL 32816, USA}

\author{Brody Yeung}
\affiliation{CREOL, The College of Optics and Photonics, University of
Central Florida, Orlando, FL 32816, USA}

\author{Benjamin Sprenger}
\affiliation{CREOL, The College of Optics and Photonics, University of
Central Florida, Orlando, FL 32816, USA}

\author{Leland Nordin}
\email{leland.nordin@ucf.edu}
\affiliation{CREOL, The College of Optics and Photonics, University of
Central Florida, Orlando, FL 32816, USA}
\affiliation{Department of Materials Science and Engineering, University
of Central Florida, Orlando, FL 32816, USA}

\date{\today}

\begin{abstract}

Detector cost remains a barrier to the widespread adoption of mid-wave infrared (3--5~\textmu m) technology. PbSe, an inexpensive narrow band gap IV--VI semiconductor that has been used since the early 1940s, delivers high performance for infrared detection despite the abundance of grain boundaries in polycrystalline films. Epitaxial growth, however, could provide superior crystalline quality and interfaces, but suitable substrates remain limited for PbSe. Recent PbSe heteroepitaxy has focused primarily on III--V, II--VI and group-IV substrates that each offer a comparatively narrow range of lattice-matched alloys for heterostructure engineering. Here we show that InP-based heteroepitaxy provides access to a broader materials platform while limiting the lattice mismatch with PbSe to approximately 4\%. We grow 151-nm thick PbSe films by molecular beam epitaxy on 200-nm thick In$_{0.53}$Ga$_{0.47}$As and In$_{0.52}$Al$_{0.48}$As buffers on (001) InP substrates. Reflection high-energy electron diffraction and X-ray diffraction show (001)-oriented rock-salt PbSe with an out-of-plane lattice constant of 6.12~\AA{} on both buffers. Photoluminescence (PL) is observed from room-temperature down to 12~K and the peak wavelength red-shifts from 3.7 to 5.0~\textmu m. Under identical measurement conditions, the room-temperature peak PL intensities from films on In$_{0.53}$Ga$_{0.47}$As and In$_{0.52}$Al$_{0.48}$As are approximately 1.9x and 1.3x that of a PbSe on GaAs substrate reference, respectively. These results establish an InP-compatible platform for integrating narrow band-gap PbSe with a broad range of ternary and quaternary III--V alloys, including, for example, structures in which In$_{0.53}$Ga$_{0.47}$As serves as both a short-wave infrared absorber and a template for PbSe growth.
\end{abstract}


\maketitle

\noindent{\small\itshape The following article has been accepted by the Journal of Vacuum Science \& Technology A. After it is published, it will be found at \href{https://doi.org/10.1116/6.0005919}{https://doi.org/10.1116/6.0005919}.}

\section{Introduction}\label{sec:intro}

The mid-wave infrared (MWIR, 3--5~\textmu m) spectral range is used in applications including, but not limited to, thermal imaging, infrared astronomy, industrial process monitoring, and healthcare \cite{rogalski2003}. Despite the valuable insights enabled by MWIR spectroscopy, its widespread adoption remains limited by the high cost of MWIR sources and detectors. State-of-the-art MWIR technologies, including HgCdTe (MCT) detectors \cite{tennant2012,kopytko2022law19,rogalski2022scaling,zandian2023rule22}, interband cascade emitters \cite{meyer1996icl,vurgaftman2015icl}, and quantum cascade lasers \cite{faist1994,vitiello2025qcl30,scalari2024qcl30},  offer high performance but remain costly, motivating the development of lower-cost materials and architectures such as silicides, SiGeSn, and lead salts \cite{reyner2021ao}. Among the IV–VI (“lead-salt”) semiconductors, PbSe exhibits suppressed Auger recombination because its highly symmetric band structure requires an indirect, phonon-assisted process. Its Auger recombination rate is therefore predicted to be several orders of magnitude lower than those of competing narrow-bandgap materials \cite{zhang2020auger}. PbSe is also compatible with low-temperature growth \cite{meyer2023apl} and maintains efficient radiative recombination despite the high threading-dislocation densities associated with lattice-mismatched heteroepitaxy. Together, these properties enable growth on lower-cost dissimilar substrates, and PbSe has consequently been demonstrated on several non-native platforms. 

Single-crystal PbSe was initially grown heteroepitaxially on rock-salt NaCl and KCl substrates \cite{zemel1965leadsalt}, followed by BaF$_2$ \cite{hohnke1974pbse} and, ultimately, CaF$_2$-coated Si(111) \cite{wu1999,majumdar2003,li2009}. These advances extended earlier work on lead-chalcogenide sensors and lasers integrated with fluoride-buffered Si \cite{zogg2001,zogg1991sst,zogg1995ngs}. Growth on Si is attractive because of its low cost and scalability, although the large lattice and thermal-expansion mismatches require carefully engineered buffer layers. In situ surface treatments reduce point defect densities, while the (111) orientation facilitates dislocation glide during thermal cyclic annealing (TCA). BaF$_2$/CaF$_2$ buffer layers bridge the lattice and thermal-expansion differences between PbSe and Si, as demonstrated by microstructural studies of BaF$_2$/CaF$_2$/Si(111) heterostructures \cite{zogg1994prb,mathet1993}. Other more recent examples include PbSnSe on CdTe/Si(211) \cite{wang2008}, PbSe heterojunctions on vicinal Ge(100) \cite{mcdowell2022}, and monolithic PbSe/Ge/Si stacks\cite{liu2026}. These advances in heteroepitaxial growth, driven primarily by the pursuit of low-cost high-performance infrared devices, have enabled successful device demonstrations such as PbSe/CdTe single-quantum-well detectors \cite{chusnutdinow2017}, CdS/PbSe photovoltaic heterojunctions \cite{weng2014}, and epitaxial PbSe/CdSe/Bi$_2$Se$_3$ unipolar-barrier ($p$Bn) detectors \cite{su2026}.
\begin{figure*}[t]
\centering
\includegraphics[width=\textwidth]{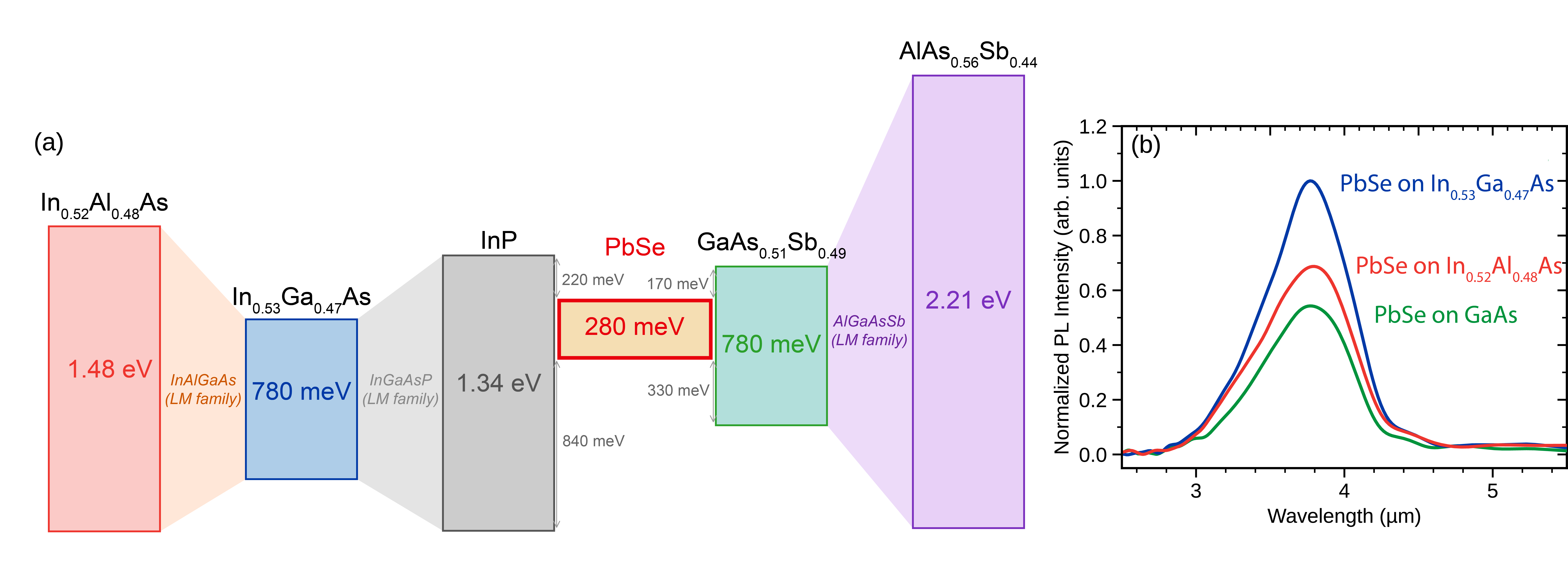}
\caption{\textbf{(a)} Room-temperature band alignment of the InP-lattice-matched III--V family against PbSe. Group-III--V band gaps are taken from the compiled parameters and bowing relations of Ref.~\onlinecite{vurgaftman2001}. The PbSe placement and associated offsets are estimates obtained from literature electron affinities following the approach used to predict the PbSe/GaAs alignment in Ref.~\onlinecite{meyer2026}. \textbf{(b)} Room-temperature mid-infrared photoluminescence of PbSe grown on the In$_{0.53}$Ga$_{0.47}$As/InP and In$_{0.52}$Al$_{0.48}$As/InP buffers compared with a PbSe/GaAs reference under the same measurement conditions. The three spectra share a common normalization to the highest-intensity PbSe on In$_{0.53}$Ga$_{0.47}$As measurement. The peak intensities of the In$_{0.53}$Ga$_{0.47}$As- and In$_{0.52}$Al$_{0.48}$As-buffered films are approximately 1.9x and 1.3x that of the GaAs-grown reference, respectively.}\label{fig:design}
\end{figure*}

III–V substrates provide additional opportunities for heterostructure engineering, and GaAs has recently emerged as the compelling III–V platform for PbSe heteroepitaxy \cite{wang2009gaas,haidet2020,haidet2023strain} and, more recently, device integration \cite{meyer2026}. Single-orientation, cube-on-cube nucleation of rock-salt PbSe on zinc-blende (001) surfaces is now well controlled \cite{haidet2020,haidet2021inas}, with the resulting films exhibiting MWIR photoluminescence (PL)\cite{meyer2021apl}, despite threading-dislocation densities on the order of $10^{9}$~cm$^{-2}$. Building on these materials advances, electrically injected $n$-PbSe/$p$-GaAs light-emitting diodes have been demonstrated\cite{meyer2026} with emission near 3.8~\textmu m and pulsed output powers of up to $\sim$400~\textmu W. Incorporating Sn into the PbSe active region extends the emission wavelength to 5~\textmu m, and these IV–VI/III–V heterojunctions have also demonstrated MWIR photodetection. However, the available lattice-matched alloys depend on the III--V substrate and GaAs primarily supports AlGaAs/GaAs heterostructures. Other III–V substrates, however, support more III–V ternary, quaternary, and quinary materials and thus a broader range of heterostructure engineering opportunities. InP, for example, is lattice-matched to In$_{0.53}$Ga$_{0.47}$As, In$_{0.52}$Al$_{0.48}$As, and related quaternaries with band gaps spanning roughly 0.7--1.5~eV. Figure \ref{fig:design}(a) compares the band alignment of PbSe with those of the InP-lattice-matched III–V material family and the wider-bandgap GaAs$_{0.51}$Sb$_{0.49}$/AlAs$_{0.56}$Sb$_{0.44}$ system accessible through quaternary alloys (the band offsets are estimated using parameters from  Ref.~\onlinecite{vurgaftman2001}). GaSb and InAs, with lattice constants of $\approx$6.1~\AA, are nearly lattice-matched to rock-salt PbSe and support alloys that include wide-gap AlAsSb barriers and type-II superlattices used in III--V infrared detectors \cite{maimon2006nbn, rogalski2017t2sl,martyniuk2014}, but do not offer the mature semiconductor ecosystem of Si, GaAs, or InP.  PbSe nucleation, interface structure, and strain relief have also been studied on GaSb and InAs substrates\cite{haidet2020,haidet2021inas,haidet2023strain}, but to our knowledge, vapor-phase heteroepitaxy on InP has not previously been reported. Previous PbSe growth on InP used aqueous electrodeposition with Cd(II)-mediated nucleation and produced twinned, partially polycrystalline (111)-oriented films\cite{beaunier2000,beaunier2001}. 
Here we examine molecular beam epitaxy (MBE) growth of PbSe on (001) InP using nominally lattice-matched In$_{0.53}$Ga$_{0.47}$As and In$_{0.52}$Al$_{0.48}$As buffers. InP ($a = 5.87$~\AA) reduces the lattice mismatch to rock-salt PbSe ($a = 6.12$~\AA) from $\sim$8\% for GaAs ($a = 5.65$~\AA) to $\sim$4\%. The ternary buffer layers maintain registry with the substrate while presenting an As-terminated zincblende surface for PbSe nucleation. We compare the resulting films by high-resolution X-ray diffraction (HRXRD) and temperature-dependent PL. Both samples show (001)-oriented rock-salt PbSe and MWIR PL at room temperature [Fig.~\ref{fig:design}(b)]. One possible use of this alloy set is a dual-band infrared detector in which In$_{0.53}$Ga$_{0.47}$As, an established short-wave infrared (SWIR) absorber with a cutoff of $\approx$1.68~\textmu m, serves as the SWIR absorber and PbSe as the MWIR absorber. In$_{0.52}$Al$_{0.48}$As or an intervening quaternary could provide an additional band offset, enabling a bias-selectable dual-color detector. The choice of blocking layer to realize this design will require a careful evaluation of the band offsets between PbSe and the III--V barrier material, as well as between In$_{0.53}$Ga$_{0.47}$As and that same barrier layer.
\begin{figure*}[t]
\centering
\includegraphics[width=\textwidth]{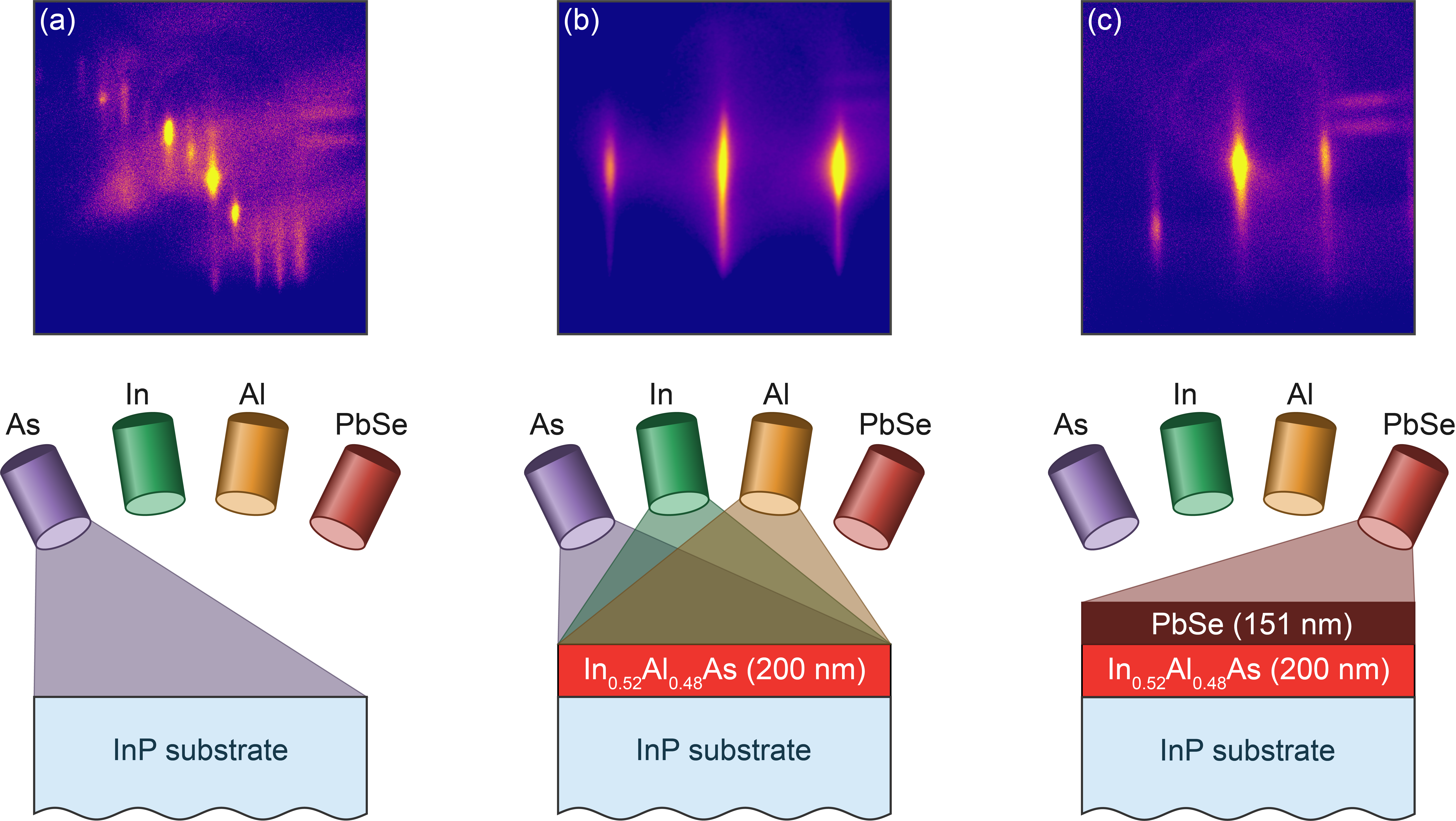}
\caption{Reflection high-energy electron diffraction (RHEED) at three stages of growth of the In$_{0.52}$Al$_{0.48}$As-buffered sample. Each panel shows the RHEED pattern together with a schematic of the corresponding source configuration. \textbf{(a)} The InP substrate surface after oxide desorption, showing the characteristic As-terminated $(2\times4)$ reconstruction, \textbf{(b)} the In$_{0.52}$Al$_{0.48}$As buffer layer, and \textbf{(c)} the PbSe film grown on the buffer. Sharp, streaky patterns are observed at every stage, consistent with two-dimensional growth and no additional RHEED features attributable to competing orientations were observed in the measured azimuth. The In$_{0.53}$Ga$_{0.47}$As-buffered sample showed similar RHEED behavior (not shown).}\label{fig:rheed}
\end{figure*}
\section{Experimental}\label{sec:methods}
All samples were grown in a MOD GenII MBE system equipped with dual-filament effusion cells for the group-III elemental sources (In, Ga, and Al), a compound PbSe source, and a Veeco Mark IV valved As cracker. Substrate temperatures were measured throughout the growth by band-edge thermometry. Epi-ready InP substrates were first deoxidized under an As overpressure while the substrate temperature was ramped to approximately 545~$^{\circ}$C. Oxide desorption was taken to be complete upon the emergence of the characteristic $(2\times4)$ InP surface reconstruction in reflection high-energy electron diffraction (RHEED), shown in Fig.~\ref{fig:rheed}(a).
The substrate was then immediately cooled to 490~$^{\circ}$C, at which $\approx$200~nm of nominally lattice-matched In$_{0.53}$Ga$_{0.47}$As or In$_{0.52}$Al$_{0.48}$As was grown, shown in Fig.~\ref{fig:rheed}(b). Following growth of the III--V buffer, the substrate was cooled under As flux. At 400~$^{\circ}$C, the As flux was interrupted and the surface was exposed to a PbSe flux for 30 seconds\cite{vishwanath2016controllable,haidet2021inas}. The substrate was then cooled to 330~$^{\circ}$C for growth of a $\approx$20~nm PbSe nucleation layer. The temperature was lowered to 300~$^{\circ}$C for the remainder of the film, giving a total PbSe thickness of $\approx$151~nm, shown in Fig.~\ref{fig:rheed}(c). This two-step growth approach was originally developed for PbSe nucleation on InAs~\cite{haidet2020} and successfully applied to GaAs~\cite{meyer2021apl}. The optimal nucleation and growth temperatures for PbSe on InP may differ, but its use across these three substrate systems suggests that it provides a useful starting point for further growth optimization. 

The crystallinity and out-of-plane orientation of the films were characterized by HRXRD on a Panalytical X'Pert PRO MRD diffractometer using Cu~K$\alpha_1$ radiation ($\lambda = 1.5406$~\AA). The incident beam was conditioned by a two-bounce monochromator, and the diffracted beam passed through a two-bounce analyzer crystal. This ``double-bounce,'' triple-axis configuration provides high angular resolution and selects the Cu~K$\alpha_1$ line. Symmetric $\omega$--$2\theta$ scans were recorded about the (004) reflection with a step size of $0.001^{\circ}$ and an angular window bracketing both the InP substrate and PbSe film (004) peaks. These scans were used to locate the PbSe (004) reflection and extract its out-of-plane lattice spacing. Symmetric (004) $\omega$ rocking curves were additionally recorded at fixed $2\theta$ about the PbSe (004) reflection for both In$_{0.53}$Ga$_{0.47}$As- and In$_{0.52}$Al$_{0.48}$As-buffered films and for a reference 151~nm-thick PbSe film grown on a GaAs substrate. These $\omega$ scans probe the mosaic-tilt contribution to the diffraction peak width (Sec.~\ref{sec:xrd}). The rocking curves were acquired in an open-detector (double-axis) geometry, with the analyzer crystal removed, for all three films. The same configuration was used for the In$_{0.53}$Ga$_{0.47}$As/In$_{0.52}$Al$_{0.48}$As-buffered films and the GaAs substrate reference so that their rocking-curve widths are directly comparable.
Skew-symmetric (224) measurements, which can be combined with symmetric rocking curves to separate mosaic tilt and twist and thereby refine estimates of dislocation density and character \cite{ayers1994}, were not acquired for these samples.

\begin{figure*}[t]
\centering
\includegraphics[width=\textwidth]{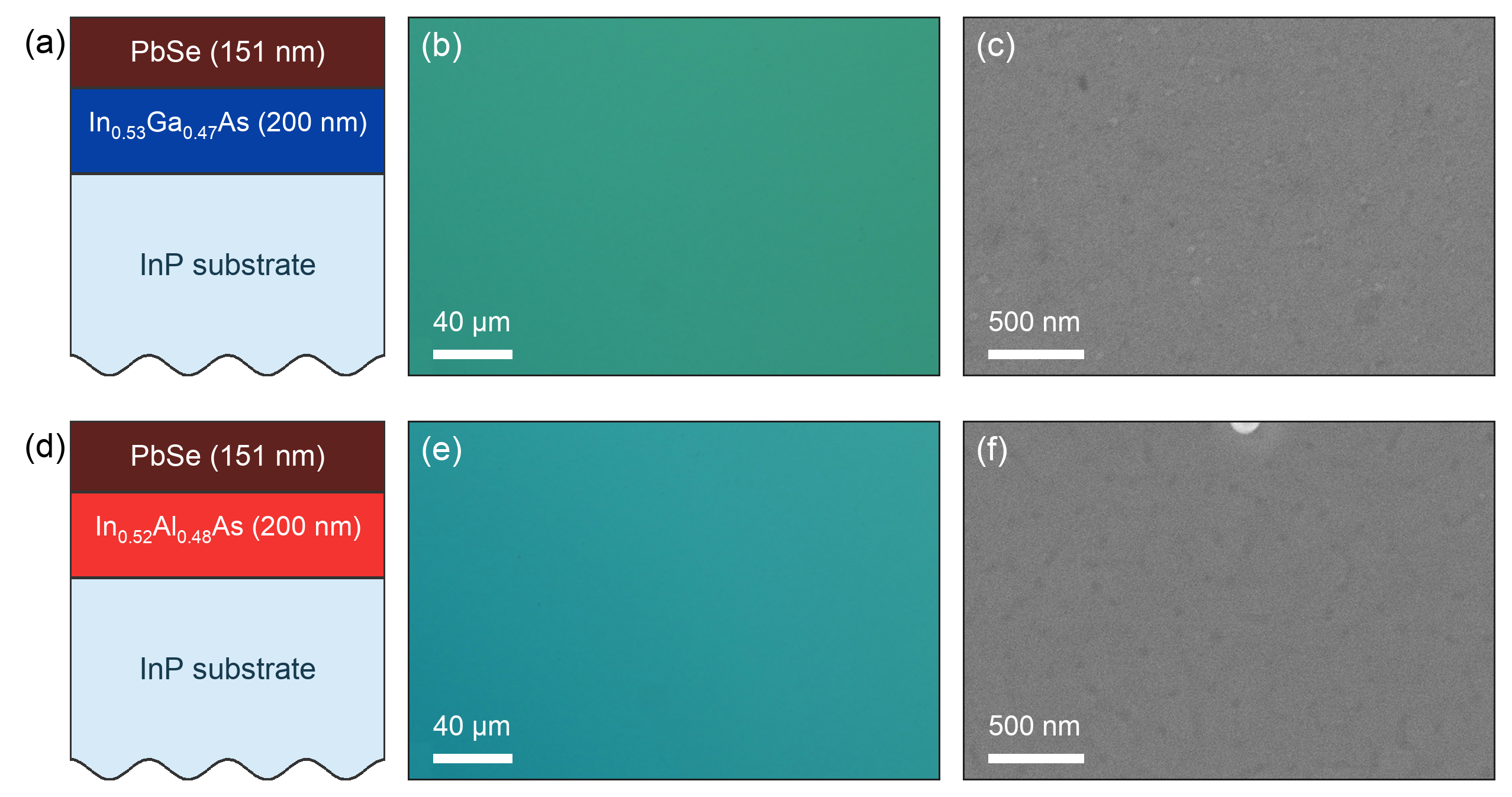}
\caption{(a,d) Schematics of PbSe films grown on (a) In$_{0.53}$Ga$_{0.47}$As/InP and (d) In$_{0.52}$Al$_{0.48}$As/InP. (b,e) Wide-field Nomarski optical micrographs and (c,f) plan-view SEM micrographs of the corresponding PbSe surfaces. PbSe films grown on both buffer layers show some surface texture in the SEM micrographs. Scale bars are 40~\textmu m in (b,e) and 500~nm in (c,f).}\label{fig:morph}
\end{figure*}

Temperature-dependent PL was measured with the samples mounted on the cold finger of a closed-cycle helium cryostat, with the temperature controlled between 12 and 300~K. The films were optically excited by a 1~W, 808~nm diode laser electrically modulated at 10~kHz, and the resulting mid-infrared emission was collected and analyzed with a Thermo Scientific Nicolet iS50 Fourier-transform infrared (FTIR) spectrometer operating in step-scan mode. A 2.4~\textmu m long-pass filter was placed before the FTIR window to block scattered pump light and isolate the mid-infrared PL, which was recorded with a liquid-nitrogen-cooled MCT detector.
For each spectrum, the photon energy axis was obtained as $E = hc/\lambda$ and the integrated intensity was computed as the baseline-subtracted trapezoidal integral over photon energy across the emission band. The band gap was estimated as $E_g = E_{\mathrm{peak}} - \tfrac{1}{2}k_B T$, where $E_{\mathrm{peak}}$ is the photon energy at the PL maximum. The nominal spectral resolution of 64~cm$^{-1}$ corresponds to an energy interval of approximately 8~meV. 
\section{Results and Discussion}\label{sec:results}

\subsection{Growth and nucleation}\label{sec:design}

The two structures, shown schematically in Figure ~\ref{fig:morph}(a,d), contain nominally 151~nm of PbSe on 200~nm In$_{0.53}$Ga$_{0.47}$As/InP or In$_{0.52}$Al$_{0.48}$As/InP. Following oxide desorption, the InP substrate exhibits the characteristic As-terminated $(2\times4)$ reconstruction [Fig.~\ref{fig:rheed}(a)]. The In$_{0.52}$Al$_{0.48}$As buffer [Fig.~\ref{fig:rheed}(b)] and the PbSe film [Fig.~\ref{fig:rheed}(c)] display sharp, streaky RHEED patterns. No additional orientations were observed. The In$_{0.53}$Ga$_{0.47}$As-buffered sample showed similar RHEED behavior at each growth stage (not shown).

Single-orientation, cube-on-cube growth of PbSe on zinc-blende III–V surfaces is thought to be controlled primarily by interfacial As bonding rather than lattice matching \cite{haidet2020}. Hence, the choice of buffer layer is guided by the surface chemistry governing rock-salt nucleation.  The lattice-matched III--V buffer, In$_{0.53}$Ga$_{0.47}$As or In$_{0.52}$Al$_{0.48}$As, supplies this As-terminated zincblende surface in registry with the InP substrate. In addition, the buffer likely acts to smooth out the epi-ready surface following thermal desorption, giving a smooth surface for the PbSe to nucleate. The directional Pb--As bonds formed at an As-terminated surface then select the cube-on-cube orientation and suppress competing variants. However, the present samples do not establish that As bonding is essential, as P could play a similar role. In contrast, attempts to prepare Sb-terminated GaSb for single-orientation PbSe nucleation using Se or PbSe flux have been reported to cause surface disordering~\cite{haidet2020}. Additional growths of PbSe on freshly grown, smooth InP buffer layers would be needed to distinguish the contributions of buffer smoothing and interfacial bonding. 

\subsection{Structural characterization}\label{sec:xrd}

Fig.~\ref{fig:morph} shows Nomarski optical micrographs (b,e) and scanning electron microscopy (SEM) images (c,f) of PbSe grown on In$_{0.53}$Ga$_{0.47}$As/InP and In$_{0.52}$Al$_{0.48}$As/InP. The Nomarski images show smooth films with no large micron-scale defects resolved at this scale. The SEM images, however, show surface texture similar to that previously observed in PbSe on GaAs following higher-temperature nucleation and subsequent growth at a slightly lower temperature~\cite{meyer2021apl}.

\begin{figure}[!t]
\centering
\includegraphics[width=\columnwidth]{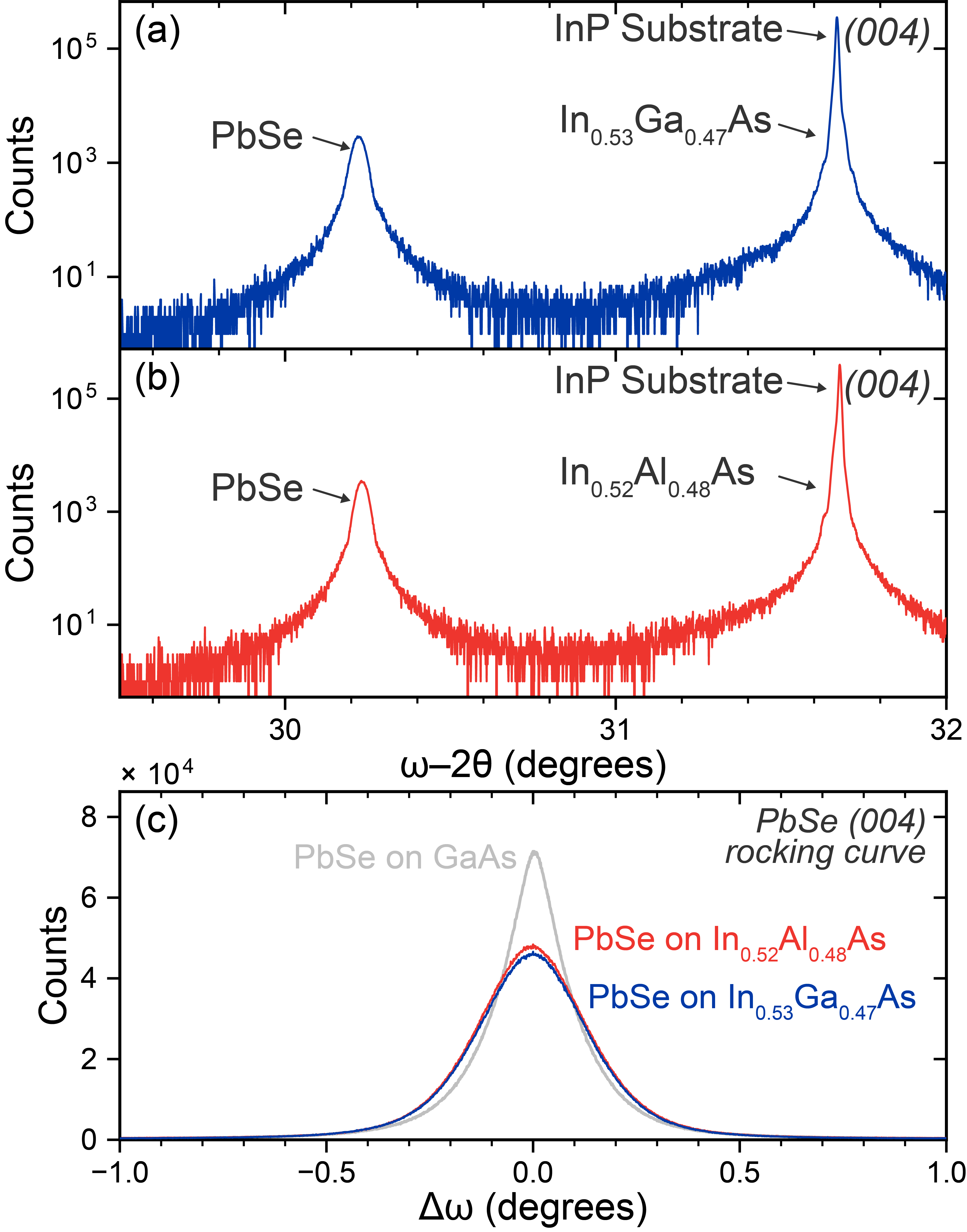}
\caption{High-resolution $\omega$--$2\theta$ X-ray diffraction scans about the (004) reflection for PbSe grown on \textbf{(a)} In$_{0.53}$Ga$_{0.47}$As/InP and \textbf{(b)} In$_{0.52}$Al$_{0.48}$As/InP. Each scan shows the InP (004) substrate peak and a PbSe (004) film reflection $\approx$1.45$^{\circ}$ lower in angle. The lattice-matched buffer layer is not separately resolved. The PbSe peak occurs at essentially the same angle for both buffers, corresponding to an out-of-plane lattice constant of $\approx$6.12~\AA. No additional PbSe reflections are observed within the scanned range. \textbf{(c)} PbSe (004) rocking curves ($\omega$ scans at fixed $2\theta$) for the same two films (colors as in panels a--b), overlaid with a rocking curve from a PbSe/GaAs reference (light gray). The open-detector rocking-curve FWHMs are 1094 and 1086~arcsec for the In$_{0.53}$Ga$_{0.47}$As- and In$_{0.52}$Al$_{0.48}$As-buffered films, respectively, and 644~arcsec for the GaAs-grown reference. Additional rocking curves about the overlapping buffer/InP substrate (004) reflection are provided in Fig.~S1 of the supplementary material.}\label{fig:xrd}
\end{figure}

High-resolution $\omega$--$2\theta$ X-ray diffraction scans about the (004) reflection are shown in Fig.~\ref{fig:xrd} for PbSe on In$_{0.53}$Ga$_{0.47}$As/InP (a) and on In$_{0.52}$Al$_{0.48}$As/InP (b). Each scan exhibits the intense InP (004) substrate reflection together with a PbSe (004) film reflection located approximately $1.45^{\circ}$ lower in angle. Because the ternary buffer layers are lattice-matched to InP, their (004) reflections overlap the substrate peak and cannot be resolved separately. Symmetric (004) $\omega$ rocking curves measured about the overlapping buffer/InP substrate reflection are provided in Fig.~S1 of the supplementary material. No additional PbSe reflections are observed within the scanned angular range. Although the symmetric scans establish the (001) out-of-plane orientation, they do not independently determine the in-plane epitaxial relationship. Nevertheless, previous studies of PbSe on GaAs and InAs suggest that an epitaxial relationship is likely \cite{haidet2020,haidet2023strain,haidet2021inas}.

The PbSe (004) reflection occurs at nearly the same angle for the two samples, corresponding to an out-of-plane lattice constant of $\approx$6.12~\AA{}, close to the bulk value of PbSe. The proximity to the bulk value suggests that the $\sim$4\% mismatch to the buffer is substantially relaxed, consistent with the range of interfacial misfit-accommodation pathways reported for PbSe on other (001) III--V templates \cite{haidet2023strain}. However, PL measurements (discussed in Sec.~\ref{sec:integ}) suggest that residual strain from coefficient of thermal expansion (CTE) mismatch is still present, as previously reported for PbSe on GaAs, a system with a similar CTE mismatch, in both absorption~\cite{liu2021strain} and luminescence~\cite{meyer2021apl}.
Within the measurement resolution, the PbSe peaks occur at the same angle for the two buffers, indicating no measurable difference in their out-of-plane lattice parameters.

The RHEED observations and symmetric diffraction are consistent with (001)-oriented PbSe on both buffers. The observed patterns are compatible with the cube-on-cube relationship commonly reported for PbSe on As-terminated zincblende surfaces, PbSe(001)$\parallel$(001) and PbSe[110]$\parallel$[110] \cite{haidet2020}, however, the present symmetric diffraction measurements do not independently establish the in-plane registry. The width of the PbSe (004) peak in the symmetric $\omega$--$2\theta$ scans provides a partial measure of crystalline quality. A pseudo-Voigt fit gives a full-width-at-half-maximum (FWHM) of $\approx$155~arcsec (0.043$^{\circ}$) for the In$_{0.53}$Ga$_{0.47}$As-buffered film and $\approx$148~arcsec (0.041$^{\circ}$) for the In$_{0.52}$Al$_{0.48}$As-buffered film. These triple-axis widths are within roughly 20\% of the $\approx$126~arcsec finite-thickness (Scherrer) limit for a 151~nm film. The coupled scans are therefore near the thickness limit and are only weakly sensitive to the mosaic structure. However, the rocking curves below probe mosaic spread more directly.

\begin{figure}[!t]
\centering
\includegraphics[width=0.96\columnwidth]{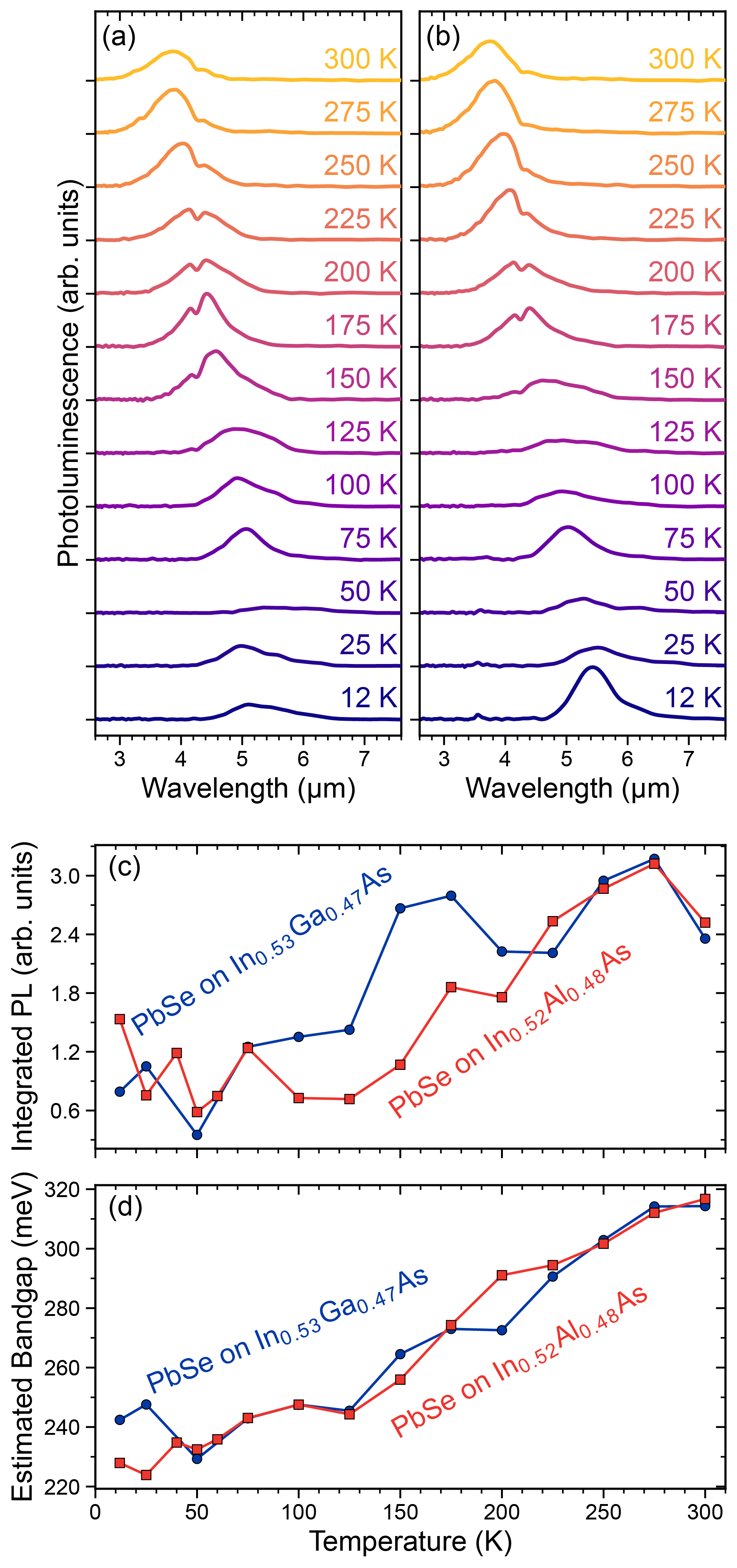}
\caption{Temperature-dependent photoluminescence (PL) of PbSe grown on \textbf{(a)} In$_{0.53}$Ga$_{0.47}$As/InP and \textbf{(b)} In$_{0.52}$Al$_{0.48}$As/InP, measured from 12 to 300~K. The waterfall spectra are vertically offset and color-coded by temperature. Each temperature series is normalized to the highest-intensity spectrum measured for the corresponding sample. \textbf{(c)} Spectrally integrated PL intensity and \textbf{(d)} estimated band gap as a function of temperature for the same two films (In$_{0.53}$Ga$_{0.47}$As/InP, blue circles and  In$_{0.52}$Al$_{0.48}$As/InP red squares). The band gap is estimated as $E_g = E_{\mathrm{peak}} - \tfrac{1}{2}k_B T$, where $E_{\mathrm{peak}}$ is the photon energy at the PL maximum. The PL remains measurable to room temperature, and $E_g$ increases overall from $\approx$230--240~meV at 12~K to $\approx$315~meV at 300~K ($\approx$+0.3~meV/K on average).}\label{fig:plwf}
\end{figure}

Symmetric (004) $\omega$ rocking curves were recorded for both In$_{0.53}$Ga$_{0.47}$As- and In$_{0.52}$Al$_{0.48}$As-buffered films and for a PbSe/GaAs reference [Fig.~\ref{fig:xrd}(c)]. Unlike the triple-axis $\omega$--$2\theta$ scans, these open-detector measurements integrate over the $2\theta$ spread and are sensitive to mosaic tilt, sub-grain curvature, and dislocation-related broadening \cite{ayers1994}. Their widths are therefore not directly comparable with the coupled-scan widths. The open-detector FWHM are 1094~arcsec (In$_{0.53}$Ga$_{0.47}$As buffer) and 1086~arcsec (In$_{0.52}$Al$_{0.48}$As buffer), compared with 155 and 148~arcsec in the triple-axis coupled scans. The difference is consistent with a mosaic-tilt contribution, but the open-detector geometry precludes quantitative separation of the broadening terms. Measured with the same rocking-curve optics, the PbSe/GaAs reference has a FWHM of 644~arcsec. It therefore has less total angular broadening than the two In$_{0.53}$Ga$_{0.47}$As/In$_{0.52}$Al$_{0.48}$As-buffered films in this measurement. Published PbSe rocking-curve widths on InAs and GaAs are also smaller \cite{haidet2020,wang2009gaas}, but differences in detector geometry and film thickness prevent direct quantitative comparison.

For an order-of-magnitude comparison with previously reported PbSe films, we convert the open-detector rocking-curve widths using the conventional mosaic-broadening relation
$N_{\mathrm{app}} \approx \beta^2/(4.36b^2)$, where $\beta$ is the FWHM in radians and $b=a/\sqrt{2}\approx4.33$~\AA\ is taken as the magnitude of a perfect PbSe dislocation Burgers vector \cite{ayers1994,meyer2023apl}. This gives apparent threading-dislocation-density scales of $N_{\mathrm{app}}\approx3.4\times10^{9}$~cm$^{-2}$ for the In$_{0.53}$Ga$_{0.47}$As-buffered film and $\approx3.4\times10^{9}$~cm$^{-2}$ for the In$_{0.52}$Al$_{0.48}$As-buffered film. The corresponding value for the GaAs-grown reference is $\approx1.2\times10^{9}$~cm$^{-2}$. These values should be regarded as upper-bound, order-of-magnitude estimates rather than measurements of a particular dislocation character because the symmetric open-detector widths contain contributions from mosaic tilt, $2\theta$ integration, microstrain, wafer curvature, and instrumental broadening. Skew-symmetric measurements would be required to separate tilt and twist and to distinguish edge, screw, and mixed dislocation contributions.

\subsection{Temperature-dependent photoluminescence}\label{sec:pl}

Temperature-dependent PL spectra are shown as waterfall plots in Figure.~\ref{fig:plwf} for the two (a) In$_{0.53}$Ga$_{0.47}$As- and (b) In$_{0.52}$Al$_{0.48}$As-buffered samples, measured from 12~K to 300~K. Both buffers produce measurable mid-infrared PL across the entire temperature range, including at room temperature. Similar persistence of PL at high threading-dislocation densities has been observed in PbSe grown on GaAs \cite{meyer2021apl,meyer2023apl,meyer2026} and is generally attributed to several factors specific to the IV--VI semiconductors. First, nonradiative recombination in PbSe is dominated by Shockley--Read--Hall processes at point defects rather than at dislocation cores, so order-of-magnitude variations in threading-dislocation density produce only modest changes in PL efficiency \cite{meyer2023apl,meyer2026}. Second, PbSe's high dielectric constant electrostatically screens charged states associated with dislocation cores, reducing their nonradiative activity \cite{wang2009gaas}. Third, PbSe's low Auger recombination coefficient limits a competing nonradiative channel at typical excitation carrier densities\cite{zhang2020auger} . 

Band alignment, however, could influence PL intensity at a IV--VI/III--V interface. PbSe on wide-gap GaAs has been reported to PL more strongly than PbSe on the more closely lattice-matched, narrow-gap InAs and GaSb, a difference attributed to carrier transfer into the III--V \cite{meyer2021apl,haidet2021inas,meyer2024adom}. Both In$_{0.53}$Ga$_{0.47}$As/In$_{0.52}$Al$_{0.48}$As-buffered films measured here emit at room temperature, despite the different In$_{0.52}$Al$_{0.48}$As (1.48~eV) and In$_{0.53}$Ga$_{0.47}$As (0.74~eV) gaps. PL intensity alone cannot determine band alignment because variations in absorption, doping, defect density, and surface recombination may also affect the measured signal. The present data therefore do not establish carrier confinement or type-I alignment. Nevertheless, the PL intensity is comparable to that of PbSe grown on GaAs. Previous work found that thin PbSe films on InAs did not exhibit detectable PbSe-characteristic PL, whereas strong room-temperature PL was observed from a much thicker, 1.5~\textmu m PbSe/InAs film \cite{haidet2021inas}. This thickness dependence suggests that the weak emission from thin PbSe/InAs films results from an interface-related carrier-loss mechanism, such as carrier transfer across an unfavorable band alignment, rather than an intrinsic inability of PbSe/InAs to PL. In contrast, the 151~nm In$_{0.53}$Ga$_{0.47}$As/In$_{0.52}$Al$_{0.48}$As-buffered films studied here emit strongly at room temperature. These results  are therefore consistent with favorable carrier confinement at the PbSe/In$_{0.53}$Ga$_{0.47}$As and PbSe/In$_{0.52}$Al$_{0.48}$As interfaces.

As a function of temperature the PL emission shifts to shorter wavelength as the temperature increases, moving from near 5~\textmu m at 12~K to $\approx$3.7~\textmu m at 300~K for both buffers. This behavior is consistent with the anomalous positive temperature coefficient of the band gap in the lead salts \cite{moss1949}, in contrast to the negative coefficient of most III--V and II--VI semiconductors. A narrow reproducible feature near 4.25~\textmu m coincides with atmospheric CO$_2$ absorption in the measurement path and is not intrinsic to the films. Importantly, the two samples show similar spectral evolution over the measured temperature range.

\subsection{Integrated intensity and band gap}\label{sec:integ}

In addition to the spectral position and relative peak magnitude of the PL, the temperature dependence of the integrated intensity provides a useful benchmark of material performance and insight into the dominant recombination processes. Fig.~\ref{fig:plwf}(c) shows the spectrally integrated PL intensity of the two In$_{0.53}$Ga$_{0.47}$As/In$_{0.52}$Al$_{0.48}$As-buffered samples. The PL remains measurable at room temperature and generally increases from 12~K to a maximum near 250–275~K, followed by a slight decrease at 300~K. Both samples exhibit a local quenching in PL intensity at 50~K that is absent at adjacent temperatures. This anomalous behavior has been reproduced in multiple temperature-dependent measurement series and in other PbSe/In$_{0.53}$Ga$_{0.47}$As samples not described here. Similar nonmonotonic low-temperature PL (i.e. not linearly increasing or decreasing with temperature) has also been reported in PbSe nanowires \cite{boercker2011}. Although the origin of the quenching in the present PbSe films remains unclear, dissociation of free excitons, followed by carrier release and nonradiative recombination, has been proposed to explain PL quenching in GaN~\cite{leroux1999quenching}. Thermally activated release of carriers from donor or acceptor states involved in donor--acceptor-pair (DAP) recombination is another possible mechanism, which has been observed in defect-luminescence quenching processes in I--VII, II--VI, and III--V compounds~\cite{reshchikov2021quenching}. The present measurements do not establish whether either mechanism accounts for the local PL quenching near 50~K. Further studies, including measurements of temperature-dependent carrier capture coefficients, will be required to identify the mechanism. The estimated bandgap [Fig.~\ref{fig:plwf}(d)] increases overall from $\approx$230–240~meV at 12~K to $\approx$315~meV at 300~K. The average positive temperature coefficient of $\approx+0.3$~meV/K is consistent with the anomalous temperature dependence of PbSe's band edge \cite{moss1949}. 

The estimated gaps exceed the accepted bulk PbSe band gap at every measured temperature. The excess is $\approx$60\% at 12~K relative to the bulk value of $\approx$0.15~eV and $\approx$15\% at 300~K relative to $\approx$0.28~eV \cite{ravindra1979pbse, preier1979}. Quantum confinement is unlikely because the 151~nm films are approximately three times thicker than the PbSe exciton Bohr radius ($\approx$46~nm) \cite{wise2000}. The systematic excess energy could be due to residual tensile strain from the mismatch in CTE between PbSe and the III--V buffer, which blue-shifts the PbSe band edge on cooling from the growth temperature and has been shown to quantitatively account for a similar bandgap increase in (001)-oriented PbSe/GaAs films in both absorption~\cite{liu2021strain} and luminescence~\cite{meyer2021apl}. In addition, a Burstein--Moss shift could be a contributing factor to the observed blueshift of the luminescence, particularly as these films were grown without excess Se and in a chamber equipped with effusion cells for In, Ga, and Al, all of which can act as $n$-type dopants for PbSe~\cite{androulakis2011pbse,evola2014pbse}. More careful measurements of the carrier concentration and strain will be required to identify each contribution to the measured PL.

\subsection{Discussion}\label{sec:discussion}
The In$_{0.53}$Ga$_{0.47}$As- and In$_{0.52}$Al$_{0.48}$As-buffered samples exhibit similar out-of-plane PbSe lattice parameters, rocking-curve widths, PL peak positions, and temperature dependence. Perhaps most importantly, under identical excitation, collection, and detection conditions, the detected room-temperature PL peak intensities of the In$_{0.53}$Ga$_{0.47}$As- and In$_{0.52}$Al$_{0.48}$As-buffered films are approximately 1.9x and 1.3x that of the PbSe/GaAs reference, respectively. Additional PbSe films grown on both buffer types under the same conditions show comparable structural and optical properties. Although the reduced nominal lattice mismatch relative to GaAs motivates the investigation of thicker PbSe layers, the present 151~nm films do not demonstrate an advantage in terms of cracking or defect density. An important next step is to incorporate this heteroepitaxial system into devices which will require direct measurements of strain, interface structure, doping, band offsets, and junction transport. In addition, systematic growth studies as functions of temperature and thickness are needed to determine the maximum PbSe thickness that can be achieved without films cracking.

\section{Summary and Conclusions}\label{sec:conclusion}

We grew PbSe by MBE on (001) InP using nominally lattice-matched In$_{0.53}$Ga$_{0.47}$As and In$_{0.52}$Al$_{0.48}$As buffer layers. RHEED and symmetric HRXRD are consistent with (001)-oriented rock-salt PbSe, and both films have an out-of-plane lattice constant near the bulk value. The (004) rocking-curve widths correspond to upper-bound, order-of-magnitude apparent threading-dislocation-density estimates in the low-$10^{9}$~cm$^{-2}$ range, above the corresponding estimate for the GaAs-grown reference measured with the same rocking-curve optics. Both In$_{0.53}$Ga$_{0.47}$As- and In$_{0.52}$Al$_{0.48}$As-buffered films exhibit PL in the mid-IR from 12 to 300~K and show the positive temperature shift characteristic of PbSe PL emission. These results demonstrate that either buffer can support PbSe growth, providing a foundation for integrating PbSe with InP-lattice-matched alloys. This platform could enable In$_{0.53}$Ga$_{0.47}$As/PbSe SWIR–MWIR heterostructures as well as more advanced detector and light-emitting-device architectures. Furthermore, the high refractive index of PbSe, combined with the lower refractive indices of InP-based materials, could provide substantial optical contrast for high-index nanophotonic structures and devices \cite{khurgin2022expanding,doiron2022supermossian}.

\section*{Supplementary Material}

See the supplementary material for symmetric (004) $\omega$ rocking curves measured about the overlapping buffer/InP substrate reflection for the PbSe/In$_{0.53}$Ga$_{0.47}$As/InP and PbSe/In$_{0.52}$Al$_{0.48}$As/InP structures (Fig.~S1).

\begin{acknowledgments}
We gratefully acknowledge support from Sandia National Laboratories and the Air Force Research Lab, through the sponsorship of an NSF I/UCRC MIST Center Project under NSF Award 1939050. Sandia National Laboratories is a multimission laboratory managed and operated by National Technology and Engineering Solutions of Sandia, LLC, a wholly owned subsidiary of Honeywell International, Inc., for the U.S. Department of Energy's National Nuclear Security Administration under contract DE-NA0003525. This paper describes objective technical results and analysis. Any subjective views or opinions that might be expressed in the paper do not necessarily represent the views of the U.S. Department of Energy or the United States Government.
This material is based upon work supported by the Air Force Office of Scientific Research under award number
FA9550-25-1-0351. Any opinions, findings, and conclusions or recommendations expressed in this material are those of the author(s)
and do not necessarily reflect the views of the United States Air Force.
\end{acknowledgments}

\section*{AUTHOR DECLARATIONS}

\subsection*{Conflicts of Interest}

The authors have no conflicts to disclose.

\subsection*{Author Contributions}
\textbf{Biridiana Rodriguez}: Formal analysis (lead); Investigation (lead); Methodology (equal); Validation (equal); Visualization (equal); Writing -- original draft (equal); Writing -- review \& editing (equal).
\textbf{Mark Martino}: Investigation (supporting); Writing -- review \& editing (supporting).
\textbf{Brody Yeung}: Investigation (supporting); Writing -- review \& editing (supporting).
\textbf{Benjamin Sprenger}: Investigation (supporting); Writing -- review \& editing (supporting).
\textbf{Leland Nordin}: Conceptualization (lead); Data curation (supporting); Formal analysis (equal); Funding acquisition (lead); Investigation (supporting); Methodology (equal); Project administration (lead); Resources (supporting); Supervision (lead); Validation (supporting); Visualization (supporting); Writing -- original draft (equal); Writing -- review \& editing (equal).
All authors reviewed and approved the final manuscript.

\section*{DATA AVAILABILITY}

The data that support the findings of this study are available from the corresponding author upon reasonable request.

\bibliographystyle{aipnum4-1}
\bibliography{sn-bibliography}

\end{document}